\documentclass[reprint,amsfonts,showpacs,pas,prl,english,superscriptaddress,float]{revtex4-1}
\usepackage{graphicx}
\usepackage{overpic}
\usepackage{color}
\usepackage{rotating}
\usepackage{txfonts}
\usepackage{array}
\usepackage{multirow}
\usepackage{url}

\draft 

\begin{document}

\title{Multi-chromatic narrow-energy-spread electron bunches from laser wakefield acceleration with dual-color lasers} 

\author{M. Zeng}
\affiliation{Key Laboratory for Laser Plasmas (Ministry of
Education), Department of Physics and Astronomy, Shanghai Jiao
Tong University, Shanghai 200240, China}
\affiliation{IFSA Collaborative Innovation Center, Shanghai Jiao Tong University, Shanghai 200240, China}
\author{M. Chen}
\email{minchen@sjtu.edu.cn}
\affiliation{Key Laboratory for Laser Plasmas (Ministry of Education), Department of Physics and Astronomy, Shanghai Jiao Tong University, Shanghai 200240, China}
\affiliation{IFSA Collaborative Innovation Center, Shanghai Jiao Tong University, Shanghai 200240, China}
\author{L. L. Yu}
\affiliation{Key Laboratory for Laser Plasmas (Ministry of
Education), Department of Physics and Astronomy, Shanghai Jiao
Tong University, Shanghai 200240, China}
\affiliation{IFSA Collaborative Innovation Center, Shanghai Jiao Tong University, Shanghai 200240, China}
\author{W. B. Mori}
\affiliation{University of California, Los Angeles, California 90095, USA}
\author{Z. M. Sheng}
\email{zmsheng@sjtu.edu.cn} \affiliation{Key Laboratory for Laser
Plasmas (Ministry of Education), Department of Physics and
Astronomy, Shanghai Jiao Tong University, Shanghai 200240, China}
\affiliation{IFSA Collaborative Innovation Center, Shanghai Jiao Tong University, Shanghai 200240, China}
\affiliation{SUPA, Department of Physics, University of
Strathclyde, Glasgow G4 0NG, UK}
\author{B. Hidding}
\affiliation{SUPA, Department of Physics, University of
Strathclyde, Glasgow G4 0NG, UK}
\author{D. Jaroszynski}
\affiliation{SUPA, Department of Physics, University of
Strathclyde, Glasgow G4 0NG, UK}
\author{J. Zhang}
\affiliation{Key Laboratory for Laser Plasmas (Ministry of
Education), Department of Physics and Astronomy,  Shanghai Jiao
Tong University, Shanghai 200240, China}
\affiliation{IFSA Collaborative Innovation Center, Shanghai Jiao Tong University, Shanghai 200240, China}

\begin{abstract}
A method based on laser wakefield acceleration with controlled
ionization injection triggered by another frequency-tripled
laser is proposed, which can produce electron bunches with low energy
spread. As two color pulses co-propagate in the
background plasma, the peak amplitude of the combined laser field
is modulated in time and space during the laser propagation due to the plasma dispersion.
Ionization injection occurs when the peak amplitude exceeds
certain threshold. The threshold is exceeded for
limited duration periodically at different propagation distances, leading to
multiple ionization injections and separated electron bunches.
The method is demonstrated through multi-dimensional
particle-in-cell simulations. Such electron bunches may be used to
generate multi-chromatic X-ray sources for a variety of
applications.
\end{abstract}

\maketitle

Versatile X-ray sources with tunable brightness, spectrum range,
pulse duration, and temporal-spatial coherence could be tools for scientific discoveries as well as medical and
industrial applications. Continuous efforts are being made to push
X-ray sources towards new limits such as coherent X-rays over
10keV~\cite{IshikawaNatPhonton2012,EmmaNatPhonton2010}, attosecond pulses~\cite{RevModPhys.81.163},
two-color hard X-ray lasers~\cite{SACLA}, etc. However, currently most of these
sources are based on conventional accelerators. More compact and
low cost X-ray devices with comparable quality are highly desired.
Recently, laser wakefield acceleration (LWFA) offers the possibility for a new generation of compact
particle accelerators~\cite{TajimaPRL1979}. Single or multiple bunches can be generated for different applications~\cite{OguchiPOP2008,CordeNatC2013,LundhPRL2013}. LWFA based compact and low cost X and
$\gamma$-ray sources also have attracted much interest~\cite{SYChenPRL2013,SCipiccia2011NatPhys,Malka2013NatPhoton,RevModPhys.85.1}.

In spite of the significant progress made in LWFA research in the past
decade~\cite{leemans2006gev,EsareyRMP2009,MalkaPOP2012,SHookerNatPh2014,wang2013quasi,NakajimaCOL2013}, it is widely
recognized that the beam quality and stability still need to be
improved considerably before its wide applications. The injection
process is a key issue of high quality beam production. There are a few interesting schemes proposed. A cold optical injection scheme is proposed to produce ultra-low energy spread beams as shown by two-dimensional (2D) simulations~\cite{DavoinePRL2009}. Besides,
ionization-induced injection in mixed gaseous targets has been
proposed and demonstrated as an attractive scheme~\cite{MinChenJAP2006,
OzPRL2007, McGuffeyPRL2010, PakPRL2010, ClaytonPRL10,hidding1,FLiPRL2013,BourgeoisPRL2013,YuPRL2014,XLXuPRSTAB2014,ChenPRSTAB2014} due to the relatively easier operation. In this
scheme a mixture of gases is chosen. In the mixture, at least one of the gas elements has a relatively
low ionization threshold such that it is effectively pre-ionized and acts as background
plasmas, while at least another has inner shells with higher ionization thresholds
 (such as nitrogen and oxygen). The laser releases these inner shell electrons at a location within the wake such that
they can be easily trapped and accelerated.
Generally, the final electron beam energy spread is related
to the effective injection length~\cite{PollockPRL2011,
JSLiuPRL2011, MChenPOP2012}, which often leads to energy spread
much larger than $1\%$, unless some techniques such as the laser
self-focusing is used~\cite{ZengPOP2014}.

\begin{figure}
\begin{tabular}{ll}
  \begin{overpic}[width=0.48\textwidth]{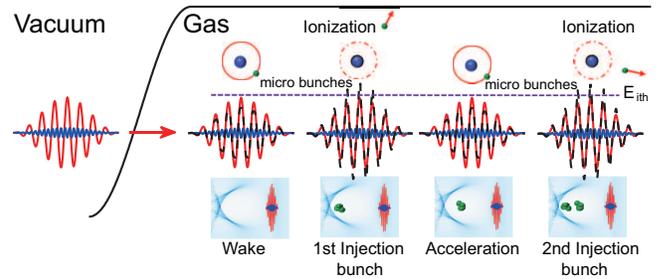}
  \end{overpic}
\end{tabular}
  \caption{\label{fig:1}(Color online) Schematic view of dual color lasers trigged periodic injection in LWFA.  A laser with base frequency $\omega_1$ (the red curves) and its harmonic $\omega_2$ (the blue curves) propagate in a mixed gas plasma. The dashed black curves show the superposition of the two frequency laser fields at
  different propagation distances. Laser parameters are chosen so that ionization-induced injection can be switched on
  when the beating is constructive and be switched off when the beating is destructive. In the plot, $E_{\rm ith}$ represents the effective threshold field for the high-Z gas inner shell ionization.}
\end{figure}

In this letter, we propose a new electron injection scheme to
produce ultralow energy spread and single or multi
electron bunches with equally spaced energy peaks. We use a
bichromatic laser to trigger sequential ionization injections. In our one-dimensional (1D) and multi dimensional particle-in-cell
(PIC) simulations using the code OSIRIS~\cite{OSIRIS}, low energy spread beams are generated because the effective injection
length is suppressed to a few hundred micrometers. Multi-peak energy spectrum can also be observed if a specific condition is satisfied.

This scenario is
illustrated in Fig.~\ref{fig:1}.  A main pulse with the fundamental
frequency is responsible for
driving an accelerating wakefield in the blowout regime~\cite{Pukhov2002,LuPRL2006,LuPRST2007}. A second co-propagating
harmonic pulse with a smaller amplitude modulates the peak
laser field strength and acts as a trigger of the high-Z gas
K-shell ionization. Due to the laser dispersion in
the plasma, the phase speed of the two frequency components are
different. By tuning the amplitude of the two components, one can
limit the K-shell ionization only occurring when the peaks of the two lasers overlap.

\begin{figure}
\begin{tabular}{ll}
  \begin{overpic}[width=0.24\textwidth]{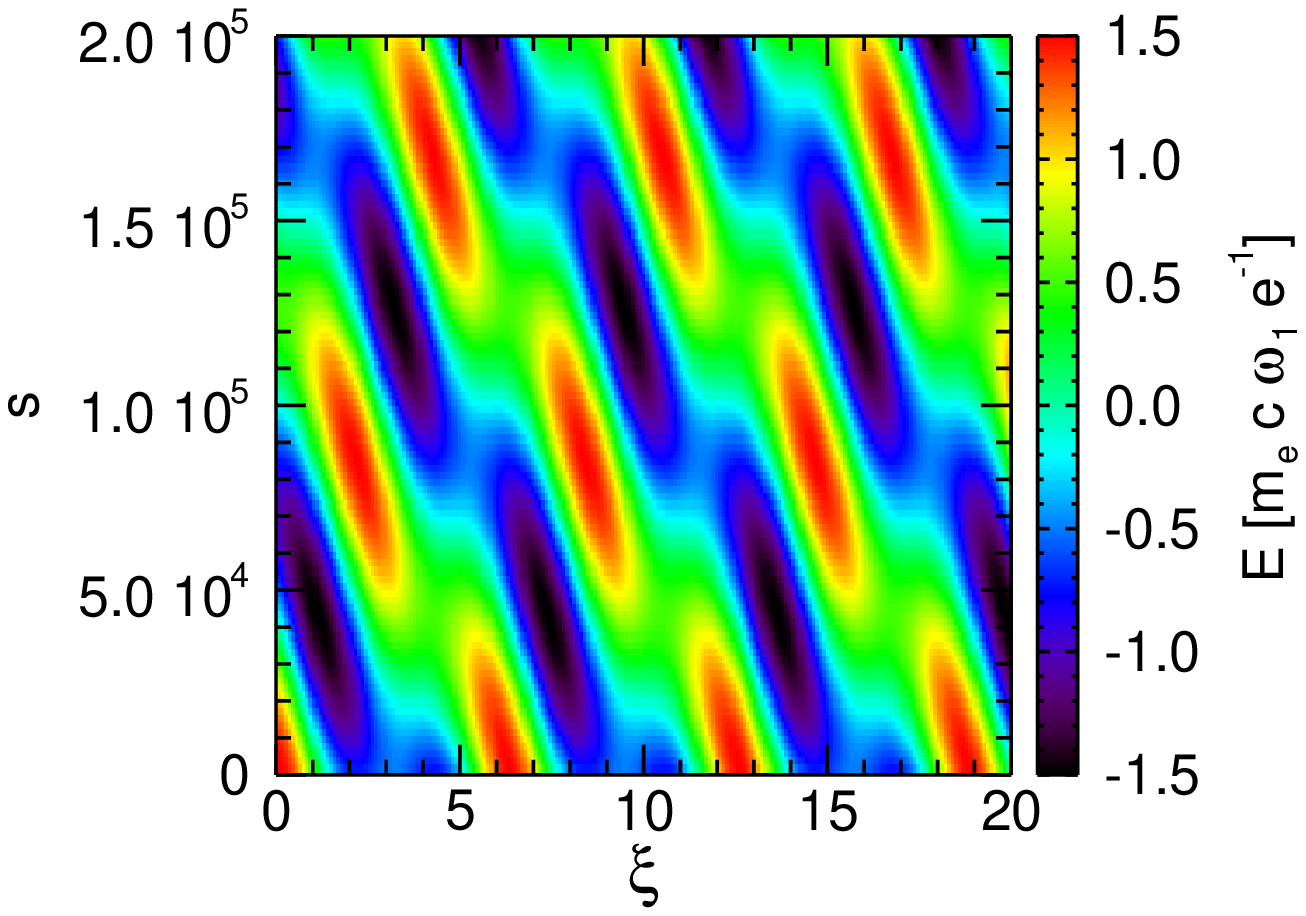}
    \put(21,62){\textcolor{white}{(a)}}
  \end{overpic}&
  \multirow{2}{*}{
  \begin{overpic}[width=0.23\textwidth, trim=18 -132 -18 132]{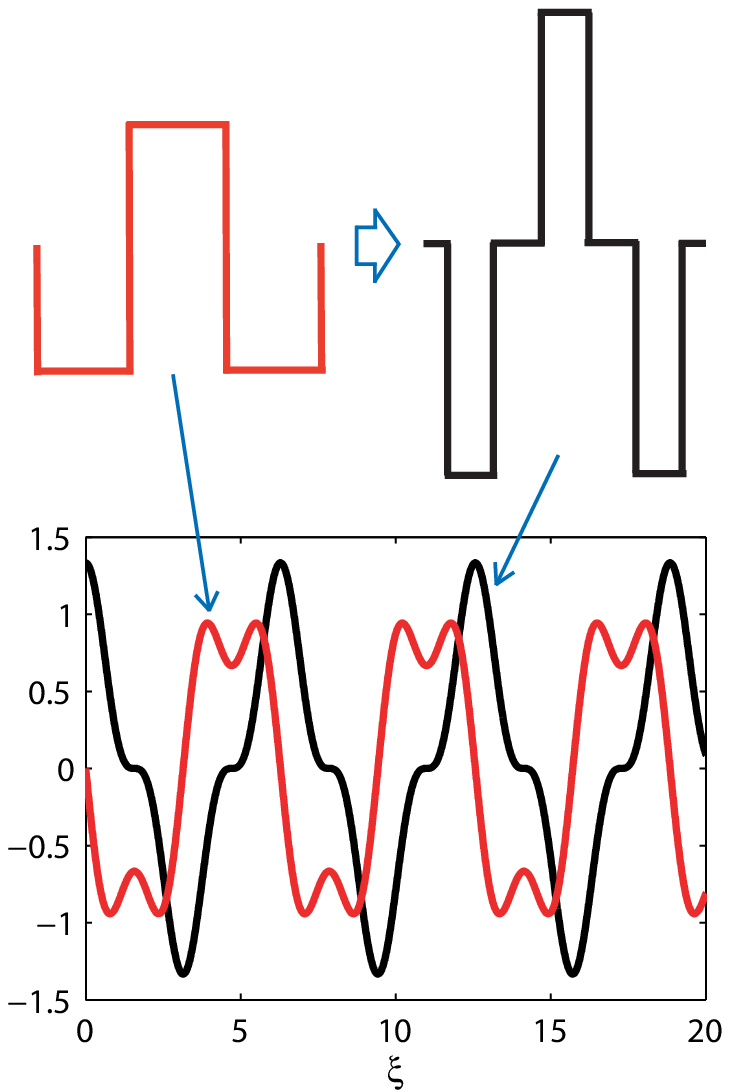}
    \put(5,140){(c)}
    \put(5,87){(d)}
  \end{overpic}} \\
  \begin{overpic}[width=0.24\textwidth]{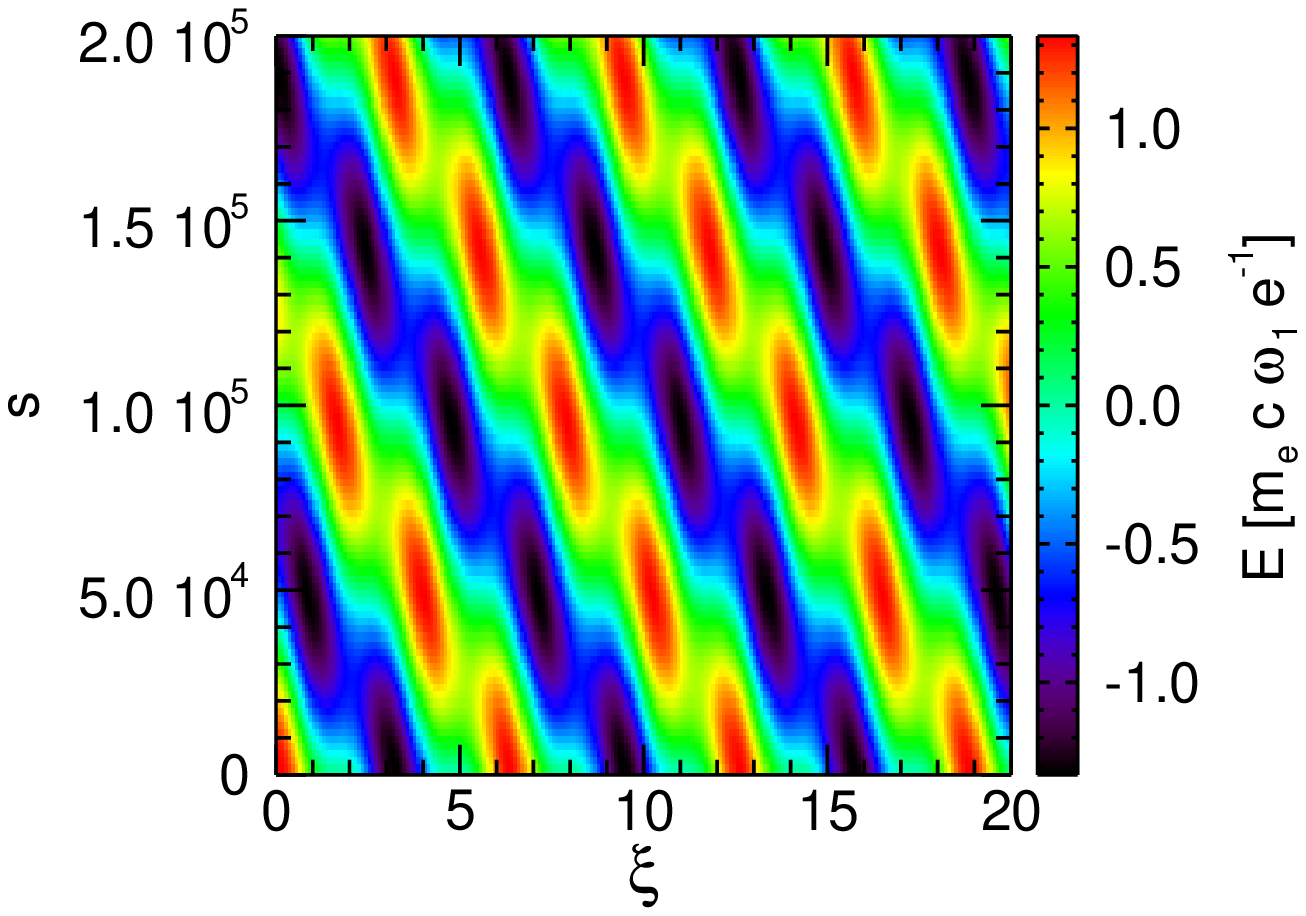}
    \put(21,62){\textcolor{white}{(b)}}
    \thicklines
    \put(21,42){\line(1,0){55}}
    \put(21,36){\textcolor{red}{\line(1,0){55}}}
  \end{overpic}&
  \multirow{2}{*}{}
\end{tabular}
  \caption{\label{fig:2}(Color online) Typical evolution of the dual color lasers according to Eq.~(\ref{eq:E_norm}).
  (a) $a_{10} = 1$, $a_{20} = 1/4$, $\omega_2 / \omega_1 = 2$ and $\omega_{\rm p}/\omega_1=0.01$. (b) $a_{10} = 1$, $a_{20} = 1/9$, $\omega_2 / \omega_1 = 3$
  and $\omega_{\rm p}/\omega_1=0.01$. (c) Illustration  showing the reason to use a SWBL combination.
  (d) Two line-outs of (b) at $s$ values indicated by the black and red lines. }
\end{figure}

To understand this process,
we first study the propagation of a bichromatic laser in
plasmas. Consider two plane waves ($i=1,2$) with the normalized vector potentials
\begin{eqnarray}\label{eq:a_TCL}
  a_i(z,t) & = & a_{i0} \sin(\omega_i t - k_i z + \phi_i),
\end{eqnarray}
where $a_{i0}$ are the amplitudes normalized to
$m_{\rm e}c^2/e$, $\omega_i$ are the frequencies, $k_i$ are the wave numbers, $\phi_i$ are the
initial phases of the two pulses, respectively. In the linear
regime, their frequencies and wave numbers satisfy the linear dispersion
relation $\omega_{i}^2=\omega_{\rm p}^2+c^2 k_{i}^2$, and
for low density plasmas the phase velocity can be expanded as ${\omega_{i} \over
k_{i}}=c(1-{1 \over 2}({\omega_{\rm p} \over \omega_{i}})^2)^{-1}$, where ${\omega_{\rm p} \over \omega_{i}} \ll 1$.
Substituting these expressions into Eq.~(\ref{eq:a_TCL}), rewriting the variables using speed of light frame variables $\xi = \omega_1 (t - {z \over c})$ and $s = {\omega_1 \over c}z$, and normalizing the frequency to $\omega_1$, time
to $\omega_1^{-1}$, length to $c/\omega_1$, the laser vector
potential can be rewritten as $a_1(\xi,s) = a_{10} \sin(\xi + {1
\over 2} \omega_{\rm p}^2 s + \phi_1)$ and $a_2(\xi,s) = a_{20}
\sin(\omega_2 \xi + {1 \over 2} {\omega_{\rm p}^2 \over \omega_2}
s + \phi_2)$. Correspondingly, the electric fields are normalized to
$m_{\rm e}\omega_1 c/e$ and can be written as
\begin{eqnarray}\label{eq:E_norm}
    E_1(\xi,s) & = & a_{10} \cos(\xi + {1 \over 2} \omega_{\rm p}^2 s + \phi_1), \nonumber \\
    E_2(\xi,s) & = & a_{20} \omega_2 \cos(\omega_2 \xi + {1 \over 2} {\omega_{\rm p}^2 \over \omega_2} s + \phi_2).
\end{eqnarray}
The total electric field is given by $E(\xi,s) = E_1(\xi,s) + E_2(\xi,s)$. We choose $\omega_2=2$, $\omega_{\rm p}=0.01$ and $a_{20}/a_{10}=1/4$ as an example, and the plot is shown in Fig.~\ref{fig:2}~(a). One can find out that changing $\phi_1$ and $\phi_2$ only leads to shifting the
pattern in Fig.~\ref{fig:2}~(a) up and down (or left and right).
The period of $s$ over which the beat pattern changes is given by
\begin{eqnarray}\label{eq:period_s}
  \Delta s = {4\pi \over \omega_{\rm p}^2(\omega_2-{1 \over \omega_2})}.
\end{eqnarray}

We only consider the situation that $a_{20} < a_{10}$ and
$\omega_2$ to be an integer larger than 1, and optimize the combination of
$a_{20} / a_{10}$ and $\omega_2$. Assume the peak field strength
of $E(\xi,s)$ for a given $s$ is $E_{\rm peak}(s)$. Its maximum
value is found at $s=s_1$ as $E_{\rm peak}|_{\rm max} = E_{\rm
peak}(s_1)$ and its minimum value is found at $s=s_2$ as $E_{\rm
peak}|_{\rm min} = E_{\rm peak}(s_2)$. Optimization for the
controlled ionization injection can be realized by tuning the ratio
\begin{eqnarray}\label{eq:ratio}
  R\left({a_{20} / a_{10}}, \omega_2\right) \equiv {E_{\rm peak}|_{\rm max} / E_{\rm peak}|_{\rm min}}.
\end{eqnarray}

It is easy to see that $E_{\rm peak}|_{\rm max} = a_{10}+a_{20}
\omega_2$, but it is not straight forward to obtain $E_{\rm
peak}|_{\rm min}$ analytically. From Eq.~(\ref{eq:E_norm}) one knows that the dispersion in
plasma does not change $\langle E^2(\xi,s)\rangle$ (the power
averaged over time $\xi$), though it changes the peak value of the
bichromatic laser field. Consider a square wave at a particular value of $s$, which has the lowest peak amplitude for a given average power. As the laser evolves the average power remains a constant, but the superposition of the laser components will become narrower peaks as shown in Fig.~\ref{fig:2}~(c). A good
approximation of a square wave is the first two components of its
Fourier series, i.~e., $\omega_2 / \omega_1=3$ and $a_{20} / a_{10}=1/9$ as shown in Fig.~\ref{fig:2}~(d). As one can see using only two frequencies approximates Fig.~\ref{fig:2}~(c) reasonably well. One can verify that the $1:3$ combination is optimal by trying other combinations and compare the ratios defined by Eq.~(\ref{eq:ratio}).
We call this combination, the square-wave like bichromatic lasers (SWBL).

\begin{figure}
\begin{tabular}{lc}
  \begin{overpic}[width=0.24\textwidth, trim=15 15 -15 -15]{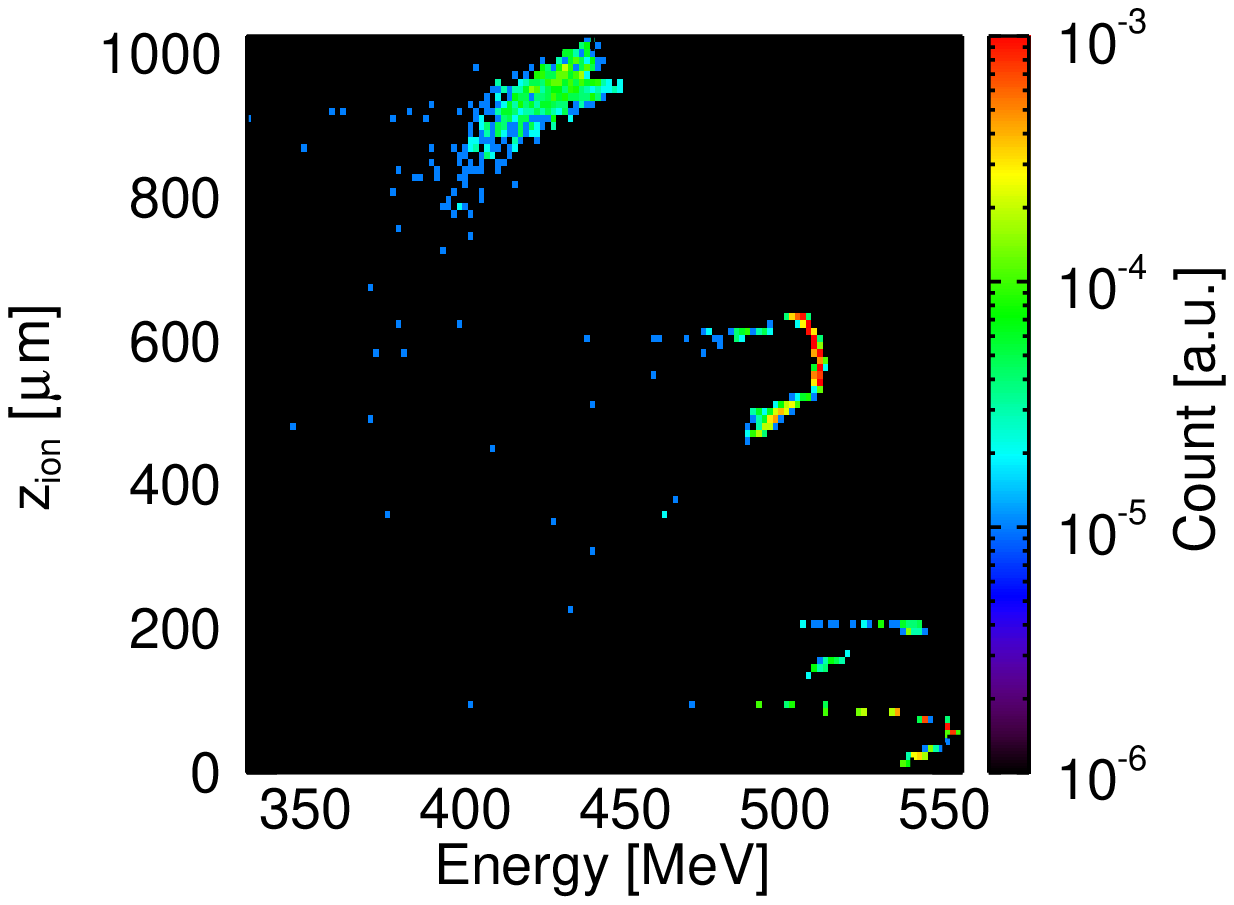}
    \put(22,61){\textcolor{white}{(a)}}
    \put(67,24){\textcolor{white}{1}}
    \put(65,41){\textcolor{white}{2}}
    \put(47,61){\textcolor{white}{3}}
  \end{overpic}&
  \begin{overpic}[width=0.23\textwidth]{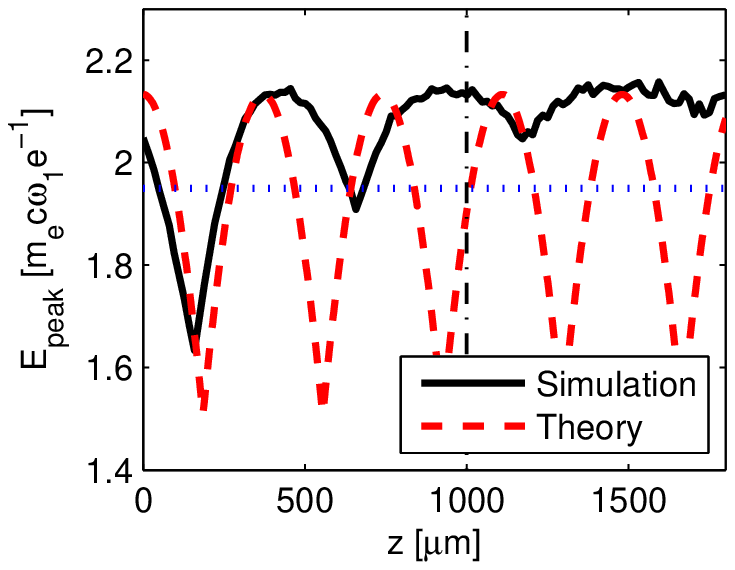}
    \put(20,14){(b)}
    \put(19,61){1}
    \put(34,61){2}
    \put(54,61){3}
  \end{overpic}\\
  \begin{overpic}[width=0.23\textwidth]{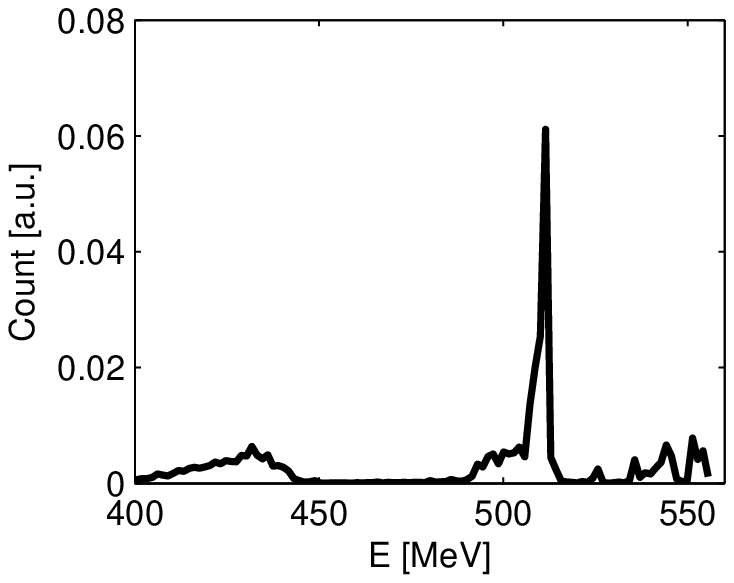}
    \put(20,62){(c)}
    \put(83,21){1}
    \put(67,60){2}
    \put(30,20){3}
  \end{overpic}&
  \begin{overpic}[width=0.24\textwidth, trim=0 5 0 -5]{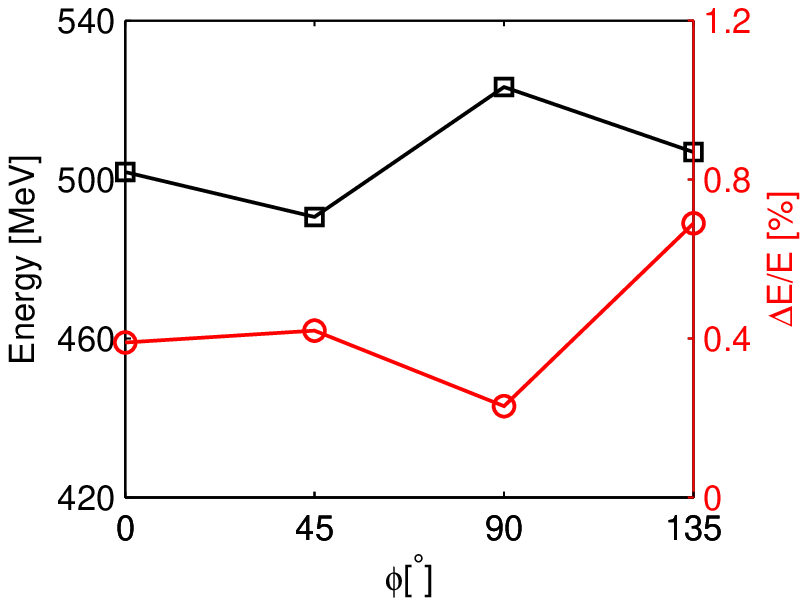}
    \put(75,63){(d)}
  \end{overpic}
\end{tabular}
  \caption{\label{fig:3}(Color online) 1D PIC simulation of the SWBL and injections. (a) Electron energy at the diagnostic point vs.\ the initial position of the injected electrons.
  (b) The SWBL peak amplitude evolution. The blue dotted line is the estimated inner shell ionization threshold, and the black dash-dotted line is the separation from the mixed gas to the pure helium gas.
  (c) Electron beam spectrum at the distance $4860\ \rm \mu m$, where the minimum energy spread during the phase rotation is measured to
  be $0.29\%$ in FWHM. (d) The energy and energy spread in FWHM vs.\ the initial laser phase $\phi$. }
\end{figure}

Based on the peculiar peak amplitude evolution of the SWBL, the
ionization injection region can be broken into small pieces. By choosing the amplitude of the SWBL so that
\begin{eqnarray}\label{eq:eth}
    E_{\rm peak}|_{\rm min} < E_{\rm N^{5+}} < E_{\rm peak}|_{\rm max},
\end{eqnarray}
the ionization injections can be limited to a few small separated
regions, where $E_{\rm N^{5+}}$ is the effective ionization threshold of $\rm N^{5+}$. One may find $E_{\rm
peak}|_{\rm max} = {4 \over 3} a_{10} m_{\rm e} c \omega_1 e^{-1}$
and $E_{\rm peak}|_{\rm min} = {2 \sqrt{2} \over 3} a_{10} m_{\rm
e} c \omega_1 e^{-1}$.

In Fig.~\ref{fig:3} we show the 1D PIC simulations results of such
multiple ionization injection and acceleration process. $\omega_1$
is chosen to be the frequency of the $800\ \rm nm$ laser and
$a_{10}=1.6$. The laser pulse duration is $33\ \rm fs$ in FWHM with
the $\sin^2$ profile. The background plasma is provided by helium
with the plasma density of $n_{\rm p}=1.6\times 10^{-3} n_{\rm
c}$, where $n_{\rm c}$ is the critical density of the $800\ \rm nm$
laser. The injection provider is nitrogen with the density of
$n_{\rm N}=1.6\times 10^{-7} n_{\rm c}$. The sequential ionization
injections can be found in Fig.~\ref{fig:3}~(a).  The curves in
Fig.~\ref{fig:3}~(b) shows the evolution of the laser peak
amplitude predicted by the theory and the result from the
simulation. The differences of the theory and the simulation after
some propagation distance are due to the plasma response and the
nonlinear laser frequency shifting~\cite{MurphyPOP2006}. To
control the injection bunch numbers we set the mixed gas length
within an appropriate length of $1\ \rm mm$. Such kind of gas jets have
already been used in several laboratories~\cite{JSLiuPRL2011}.

Totally three discrete bunches are observed in this simulation and
the energy spectrum at an  acceleration distance of $4854.4\ \rm \mu
m$ is shown in Fig.~\ref{fig:3}~(c) with the three peaks labeled,
corresponding to the three injections shown in
Fig.~\ref{fig:3}~(a) and Fig.~\ref{fig:3}~(b) within the mixed gas region. Each
injection duration is limited to $100 \sim$ $200\ \rm \mu m$. In
this specific simulation, the second bunch has its minimal energy
spread of $0.29\%$ at distance of $4860\ \rm \mu m$, while the other
two bunches can also get their minimum energy spreads at other
appropriate acceleration distances. It is worth noting that even
though the optimal acceleration distances for the minimal energy
spreads are different for different bunches, this proposed scheme
is robust because the accelerated beams can keep a low energy
spread in a sufficient wide range of acceleration distances, which
is clearer in multi-dimensional cases.

For such two color beat wave ionization injection scheme, the
initial phases of the two pulses
may affect the specific ionization
injection position. Without loss of generality, let $\phi_1 =
\phi_2 \equiv \phi$, and change $\phi$ from $0^{\circ}$ to
$135^{\circ}$, which correspondingly changes the positions of the
electric field reaching $E_{\rm peak}|_{\rm max}$. The output beam
energies and energy spreads at acceleration distance of $4800\ \rm
\mu m$ are shown in Fig.~\ref{fig:3}~(d). One can see that the
central energy of a specific electron bunch at a fixed laser
propagation distance has a fluctuation of 30MeV, which is close to
the energy difference between the first and second bunch
shown in Fig.~\ref{fig:3}~(c). This energy fluctuation
comes from the different injection positions set by the two laser
phases. Nevertheless, we find all these simulations show quite
good beam quality. They keep small energy spreads of less than
$1\%$ regardless of the initial phase change.

\begin{figure}
\begin{tabular}{lc}
  \begin{overpic}[width=0.24\textwidth]{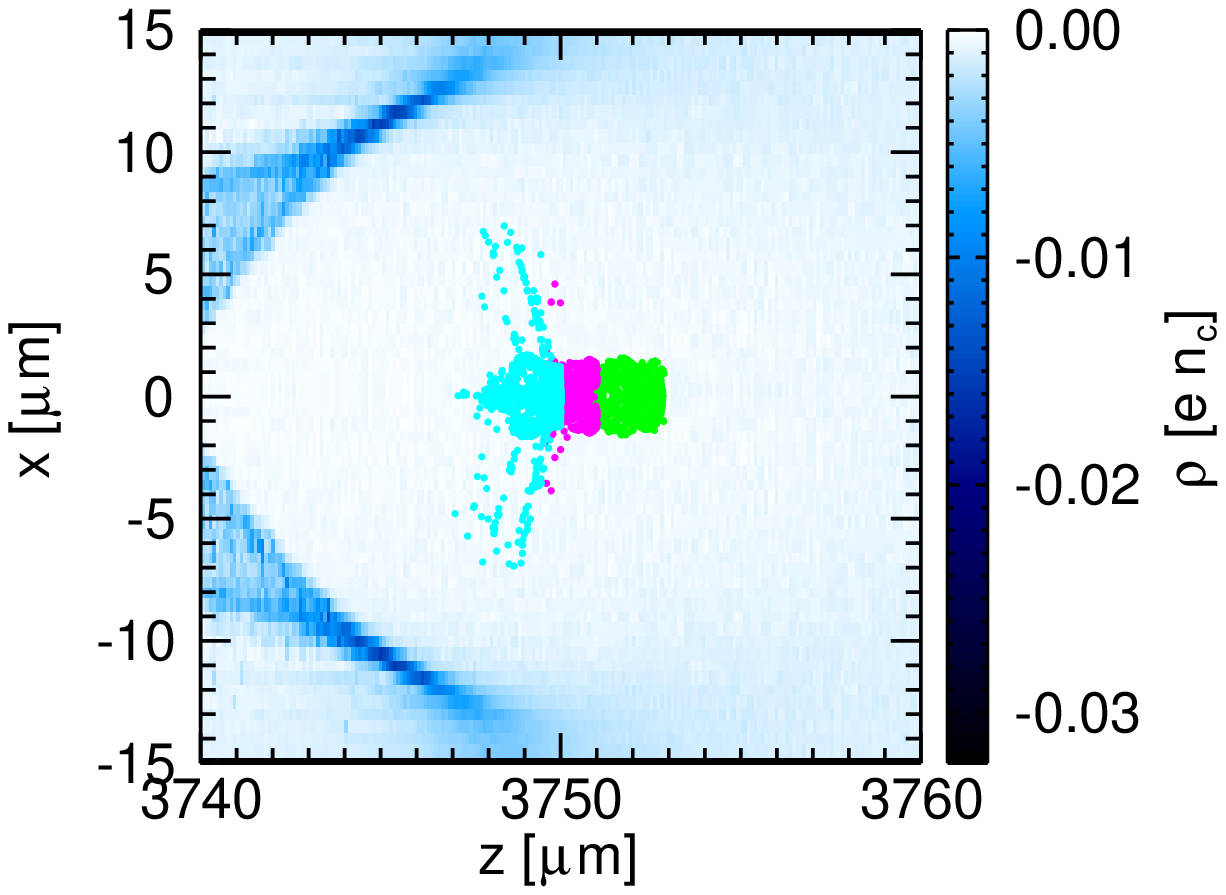}
    \put(25,62){(a)}
    \put(53,50){1}
    \put(48,50){2}
    \put(42,50){3}
  \end{overpic}&
  \begin{overpic}[width=0.24\textwidth]{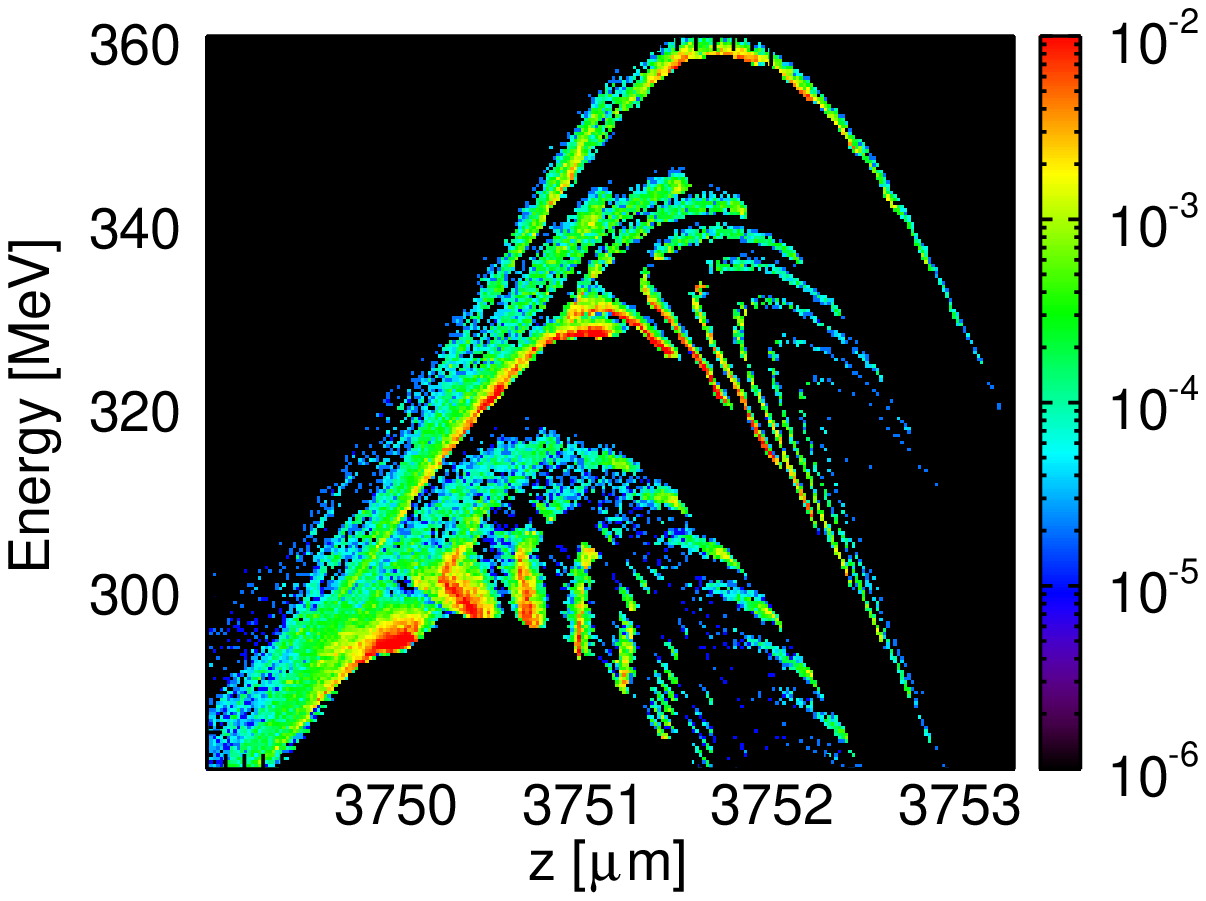}
    \put(22,62){\textcolor{white}{(b)}}
    \put(34,18){\textcolor{white}{3}}
    \put(48,43){\textcolor{white}{2}}
    \put(57,64){\textcolor{white}{1}}
  \end{overpic}\\
  \begin{overpic}[width=0.23\textwidth]{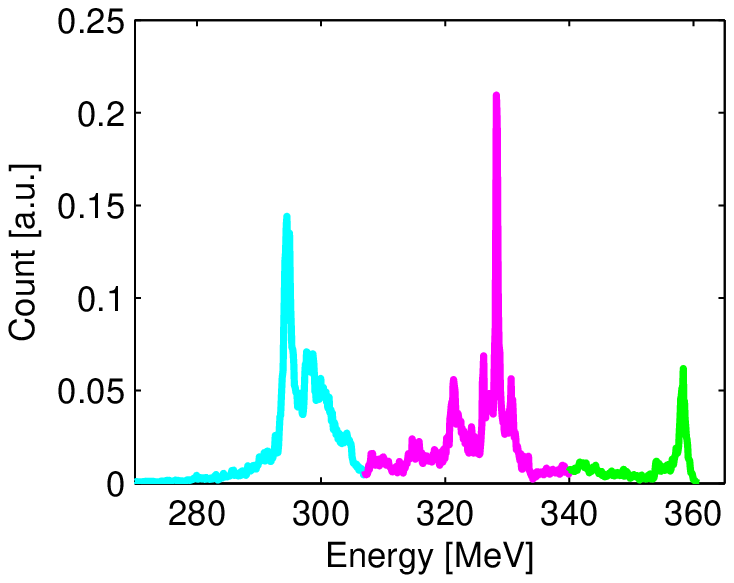}
    \put(79,61){(c)}
    \put(83,30){1}
    \put(60,63){2}
    \put(34,50){3}
  \end{overpic}&
  \begin{overpic}[width=0.23\textwidth]{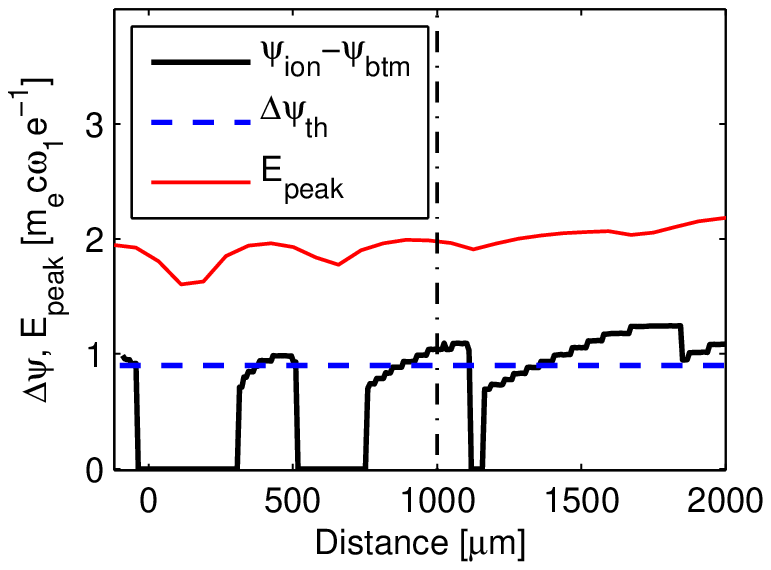}
    \put(78,60){(d)}
  \end{overpic}
\end{tabular}
  \caption{\label{fig:4}(Color online) 2D PIC simulations of the SWBL injection scheme. (a) The density snapshot at
  $z=3780\ \rm \mu m$. The colored dots show the locations of three electron bunches. (b) The energy and space distribution of
  the energetic electrons. (c) The spectrum of the injected electrons, showing three monoenergetic peaks. (d) The pseudo-potential ($\psi$) difference of the wake and the laser peak field evolution. The dash-dotted line is the separation from the mixed gas region to the pure helium region. }
\end{figure}

Beside the phase effects, other high dimensional effects such as the self-focusing, the evolution of the bubble radius, and off axis ionizations, may also affect the injection and acceleration. Among them, self-focusing and the associated evolution of the bubble radius are most important. One of the solutions is to choose a matched spot size~\cite{LuPRST2007}. Another solution is to choose a relative
large spot size so that self-focusing occurs after a sufficient
long acceleration distance, during which multiple injections have
already established. Generally, self-focusing occurs in a
distance estimated by $z_{\rm sf} = Z_{\rm R}(\frac{\alpha}{32}a_{10}^2 k_{\rm p}^2 W_0^2-1)^{-1/2}$, where $\alpha =
\sqrt{2}$ for a 2D slab geometry and $\alpha = 1$ for a 2D
cylindrical or 3D geometry~\cite{ZengPOP2014,TzengPRL1998}. With the presence
of self-focusing, a two-stage acceleration process should be
deployed~\cite{PollockPRL2011,JSLiuPRL2011}. For the
multiple-injection to occur, it is required that the injection
stage length satisfies
\begin{eqnarray}\label{eq:length_requir}
  L_{\rm inj} < z_{\rm sf}.
\end{eqnarray}
In this case, the number of ionization injected bunches can be estimated as
\begin{eqnarray}\label{eq:N_peak}
  N_{\rm bunch}  =  \left[ L_{\rm inj} / ({c \Delta s \over \omega_1}) \right],
\end{eqnarray}
where the square brackets pair means the downward rounding. The
energy difference between the monoenergetic peaks can  be
estimated by the injection position difference times the averaged
acceleration gradient
\begin{eqnarray}\label{eq:E_diff}
  \Delta {\rm Energy} =  {c \Delta s \over \omega_1} \times {1 \over 2} G_0,
\end{eqnarray}
where $G_0 [{\rm eV / m}] \approx 96 \sqrt{n_{\rm p}[{\rm cm}^{-3}]}$.

A typical 2D simulation is shown in Fig.~\ref{fig:4}, where we
choose $a_{10} = 1.46$, $a_{20} = 0.162$, $W_0 = 80\ \rm \mu m$ and other parameters are
the same as those in the 1D simulations. The initial laser amplitude in the 2D simulation is lower than that in 1D. But when the SWBL field reaches its first maximum in 2D, the peak field strength is very close to that in the 1D cases due to the self-focusing effect. The injection stage
length is $L_{\rm inj} = 1\ \rm mm$ so that
Eq.~(\ref{eq:length_requir}) is satisfied. A typical distribution of
the injected bunches is shown in Fig.~\ref{fig:4}~(a).  The central
positions of these bunches are spatially separated at this snapshot
with $\rm \mu m$ scale separations. Figure \ref{fig:4}~(b) shows the
phase space distribution of the bunches, from which we see within
the second and the third bunches, there are a few micro
bunches. These micro bunches come from the several overlapping peaks of the combined electric fields larger than the ionization threshold as schematically
shown in Fig.~\ref{fig:1}. These bunches degrade the
monochromaticity of the final beams, showing the pedestals between
the peaks of the energy spectrum in Fig.~\ref{fig:4}~(c). In
addition, the whole spectrum is composed of three main
peaks with the separation of $30\ \rm MeV$ confirming the prediction
of Eqs.~(\ref{eq:N_peak}) and (\ref{eq:E_diff}). From our
simulations we find these pedestals can be reduced by using a
shorter $3\omega$ laser, which makes the inner shell ionization
only occurs in a single overlapping electric field peak. A
simulation with $10\ \rm fs$ $3\omega$ laser gives a single injected
electron bunch with final energy spread less than $0.2\%$ in FWHM.

The injection positions of the electrons can be estimated by evaluating both the ionization threshold and the pseudo-potential ($\psi$) differences~\cite{PakPRL2010} related to the wake and the ionized electrons. The threshold for ionization injection is given by $\Delta \psi_{\rm th}=1-{\sqrt{1+(p_{\perp}/m_e c)^2} \over \gamma_{\rm ph}} \approx 0.9$, where the normalized transverse momentum is estimated to be the normalized laser vector potential at ionization $p_{\perp}/m_e c \approx 1.9$, and the wake phase velocity Lorentz factor is estimated by the linear theory $\gamma_{\rm ph} \approx \omega/\omega_{\rm p} = 25$. In Fig.~\ref{fig:4}~(d), the blue dashed line shows this threshold, and the black line shows the drop of $\psi$ from the nitrogen K-shell ionization position to the minimum $\psi$, which is manually set to zero when the laser amplitude is lower than the K-shell ionization threshold. There are three periods in the injector region (distance up to $1000\ \rm \mu m$) that satisfy the injection condition, consist with the three period when laser peak field exceeds the effective ionization threshold of the nitrogen inner shell, and also consist with the three injected bunches. In the simulation, we found that within a larger distance (between $3.5\ \rm mm$ and $4.5\ \rm mm$), all of the bunches keep very low energy spread (less than $0.4\%$ in FWHM). This gives a very larger acceleration distance window to get a high quality beam in experiments.

\begin{figure}
\begin{tabular}{lc}
  \begin{overpic}[width=0.24\textwidth,  trim=40 0 10 0]{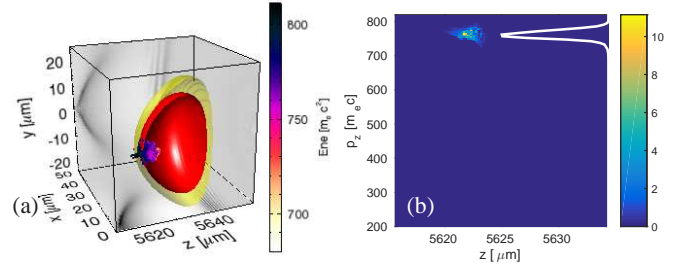}
    \put(0,15){(a)}
  \end{overpic}
  \begin{overpic}[width=0.24\textwidth]{TCL3D_12_p1x1_450.eps}
    \put(20,15){\textcolor{white}{(b)}}
  \end{overpic}
\end{tabular}
  \caption{\label{fig:5}(Color online) A 3D PIC simulation of the SWBL injections. (a) A 3D wake plot. Only a half of the bubble is plotted to show the inner structure of the bubble and the injected electrons. The electron beam has normalized emittances of $3.3\ \rm \mu m \cdot rad$ in the laser polarization direction, and $2.3\ \rm \mu m \cdot rad$ in the perpendicular direction. (b) Phase space of the injected charge. The white curve is the projection to the $p_z$ axis. }
\end{figure}

A series of 3D simulations are also performed. A typical result is shown in Fig.~\ref{fig:5}, in which we choose $n_{\rm p} = 8 \times 10 ^{-4} n_{\rm c}$, $a_{10}=1.485$, $W_0=40\ \rm \mu m$ and $L_{\rm inj}=1\ \rm mm$ so that $N_{\rm bunch} = 1$ according to Eq.~(\ref{eq:N_peak}). A beam with a total charge of $12.6\ \rm pC$, a mean energy of $389 \rm MeV$ and a true RMS energy spread of $1.53\%$ is produced, which confirms the effectiveness of SWBL injection scheme. From other 3D simulations we notice that as $n_{\rm p}$ decreases (laser power should be no less than the critical power for self-guiding while keeping $a_0$ unchanged, thus $W_0$ may be increased accordingly), the laser can be self-guided longer, the final electron beam energy increases, and the relative energy spread decreases. Although we have not yet tested the GeV level acceleration due to the limited computational resources available, from the serial 3D runs with absolute energy spread $\sim 5\ \rm MeV$ it is very promising that our injection scheme can produce electron beams with energy spread lower than $1\%$ once the plasma density and laser power are suitable for GeV level accelerations.

In conclusion, we have proposed a dual color laser scheme to
control ionization injection in LWFAs. It can result in periodic
triggering of the ionization injection and consequently produce a
unique comb-like energy spectrum. These features are demonstrated
by multi-dimensional PIC simulations. The energy spread of an individual electron bunch produced from a single injection period can be controlled down to around $1\%$ or even less with the central energy of a few hundred MeV. Our scheme to generate multi-chromatic narrow energy-spread electron bunches can be used for multi color
X-ray generation~\cite{SACLA,MulticolorXRayPRL2013}, which is particularly interesting for
medical imaging applications~\cite{multi-energy-X1,
dierickx2009diagnostic}. The multi-chromatic beams may also be interesting for radiotherapy~\cite{MalkaNP2008}.

\acknowledgments{This work was supported by the National
Basic Research Program of China (Grant No.\ 2013CBA01504), the
National Natural Science Foundation of China (Grant Nos.\ 11421064,
11374209, 11374210 and 11405107), the MOST international
collaboration project (Grant No.\ 2014DFG02330), the US DOE DE-SC 0008491, DE SC 0008316, DE FG02 92ER401727 and the US NSF ACI 1339893.
M.C. appreciates supports from National 1000 Youth Talent Project of China. The authors would like to acknowledge the OSIRIS Consortium,
consisting of UCLA and IST (Lisbon, Portugal) for the use of
OSIRIS and the visXD framework. Simulations were carried
out on the $\Pi$ supercomputer at Shanghai Jiao Tong University.}

%

\end{document}